\documentclass[conference,10pt]{IEEEtran}

\usepackage{graphicx} 

\usepackage{cite}
\usepackage{amsmath,amssymb,amsfonts}
\usepackage{algorithmic}
\usepackage{graphicx}
\usepackage{textcomp}
\usepackage{xcolor}
\def\BibTeX{{\rm B\kern-.05em{\sc i\kern-.025em b}\kern-.08em
    T\kern-.1667em\lower.7ex\hbox{E}\kern-.125emX}}
\usepackage{graphicx}
\usepackage{subcaption}
\graphicspath{ {images/} }
\usepackage[utf8]{inputenc}
\usepackage{fancyhdr}
\usepackage{dirtytalk}
\usepackage{tabularx}
\usepackage[ruled,vlined]{algorithm2e}

\usepackage{enumitem,kantlipsum}
\usepackage{listings}
\usepackage{graphicx}
\graphicspath{ {images/} }
\usepackage[utf8]{inputenc}
\usepackage{fancyhdr}
\usepackage{dirtytalk}
\usepackage{tikz}
\usepackage{bbm}
\usetikzlibrary{shapes.geometric, arrows}
\usepackage{pgfplots}
\pgfplotsset{width=8cm,compat=1.16}
\usepgfplotslibrary{external}
\usepackage{array}
\newcolumntype{P}[1]{>{\centering\arraybackslash}p{#1}}

\usepackage[most]{tcolorbox}

\tikzstyle{startstop} = [rectangle, rounded corners, minimum width=2cm, minimum height=0.5cm,text centered, draw=black, fill=red!30]
\tikzstyle{process} = [rectangle, minimum width=2cm, minimum height=0.5cm, text centered, draw=black, fill=orange!30, align=left]
\tikzstyle{decision} = [diamond, minimum width=1.0cm, minimum height=0.4cm, text centered, draw=black, fill=green!30]
\tikzstyle{arrow} = [thick,->,>=stealth]

\newcolumntype{L}{>{$}l<{$}}

\usepackage[labelformat=simple]{subcaption}

\usepackage[nolist]{acronym}
\begin{acronym}
\acro{LM}{language model}
\acro{LLM}{large language model}
\acro{SLM}{small language model}
\acro{RAG}{retrieval augmented generation}
\acro{LM}{language model}
\acro{LoRA}{low-rank adaptation}
\acro{QLoRA}{quantized low-rank adaptation}
\acro{SBERT}{sentence BERT}
\acro{ML}{machine learning}
\acro{TF-IDF}{term frequency and inverse document frequency}
\acro{OOD}{out-of-distribution}
\acro{QnA}{questions and answering}
\acro{AI}{artificial intelligence}

\acro{3GPP}{third generation partnership project}
\acro{O-RAN}{open radio access networks}
\acro{MCQ}{multiple choice question}

\acro{MAS}{multi-agent system}
\acro{CRN}{cognitive radio network}

\acro{MDP}{Markov decision process}
\acro{IC}{incentive compatibility}
\acro{IR}{individual rationality}
\acro{URLLC}{ultra reliable low latency}
\acro{AI}{artificial intelligence}
\acro{SNR}{signal-to-noise-ratio}
\acro{CI/CD}{continuous integration and continuous deployment}
\acro{GAN}{generative adversarial network}
\acro{NFV}{network function virtualization}
\acro{SDN}{software-defined networking}
\acro{RLHF}{Reinforcement learning from human feedback}
\acro{QoS}{quality-of-service}
\acro{SINR}{signal-to-interference-plus-noise ratio}
\acro{SBS}{small base station}
\acro{MBS}{macro base station}

\acro{VCG}{Vickrey–Clarke–Groves}
\acro{GFP}{Generalized First Price}
\acro{GSP}{Generalized Second Price}

\acro{HetNet}{heterogeneous network}
\acro{UE}{user equipment}
\acro{BS}{base station}

\end{acronym}

\begin{document}
\pagenumbering{gobble} 
\captionsetup[figure]{labelfont={bf},labelformat={default},labelsep=period,name={Figure}}






\title{Large Language Models as Bidding Agents in Repeated HetNet Auction}

\author{\IEEEauthorblockN{Ismail Lotfi\IEEEauthorrefmark{1},
Ali Ghrayeb\IEEEauthorrefmark{1},
Samson Lasaulce\IEEEauthorrefmark{2}, and
Mérouane~Debbah\IEEEauthorrefmark{3}}

\IEEEauthorblockA{\IEEEauthorrefmark{1}College of Science and Engineering, Hamad Bin Khalifa University, Doha, Qatar}

\IEEEauthorblockA{\IEEEauthorrefmark{2}CNRS, Universite de Lorraine, CRAN, France}

\IEEEauthorblockA{\IEEEauthorrefmark{3}Department of Computer Science, Khalifa University, Abu Dhabi, UAE}

(e-mail: \{ilotfi, aghrayeb\}@hbku.edu.qa, samson.lasaulce@univ-lorraine.fr, merouane.debbah@ku.ac.ae)
}







\maketitle

\begin{abstract}
This paper investigates the integration of large language models (LLMs) as reasoning agents in repeated spectrum auctions within heterogeneous networks (HetNets). 
While auction-based mechanisms have been widely employed for efficient resource allocation, most prior works assume one-shot auctions, static bidder behavior, and idealized conditions.
In contrast to traditional formulations where base station (BS) association and power allocation are centrally optimized, we propose a distributed auction-based framework in which each BS independently conducts its own multi-channel auction, and user equipments (UEs) strategically decide both their association and bid values.
Within this setting, UEs operate under budget constraints and repeated interactions, transforming resource allocation into a long-term economic decision rather than a one-shot optimization problem. 
The proposed framework enables the evaluation of diverse bidding behaviors—from classical myopic and greedy policies to LLM-based agents capable of reasoning over historical outcomes, anticipating competition, and adapting their bidding strategy across episodes.
Simulation results reveal that the LLM-empowered UE consistently achieves higher channel access frequency and improved budget efficiency compared to benchmarks. These findings highlight the potential of reasoning-enabled agents in future decentralized wireless networks markets and pave the way for lightweight, edge-deployable LLMs to support intelligent resource allocation in next-generation HetNets.

\end{abstract}


\section{Introduction}
\subsection{Background}

Dynamic spectrum allocation in mobile networks is a critical and ongoing challenge, particularly in \acp{HetNet} where macro and small cells coexist to support diverse user demands and service requirements~\cite{Yang_2013_COMST_AuctionTheory}. Traditional allocation techniques include fixed or semi-static frequency assignments, centralized optimization algorithms, and distributed load-balancing schemes. However, the growing complexity and dynamic nature of wireless environments have led to the emergence of auction-based mechanisms as promising solutions, particularly when demands exceed the available spectrum~\cite{Yang_2013_COMST_AuctionTheory, Ismail_2021_Globecomm}. These mechanisms leverage economic principles to incentivize users to reveal their true preferences and allocate limited spectrum resources efficiently. 

Several auction-based mechanisms have been proposed in the literature for resource allocation in \acp{HetNet}, with the majority focusing on single-round (one-shot) auctions where users submit bids and receive allocations in a single instance~\cite{Dong_INFOCOM_2014, Alibraheemi_ACCESS_2024}. These models typically assume non-strategic users who truthfully report their valuations, often equated with their achievable channel capacity under ideal conditions. While such formulations simplify the analysis, they overlook the inherently dynamic nature of real-world wireless networks. In practical HetNet deployments -especially in mobile or dense urban scenarios- resource allocation decisions must be revisited frequently due to the rapid fluctuations in user demands, channel conditions, and network loads. This naturally gives rise to repeated auction settings, where spectrum resources are re-allocated periodically~\cite{Mehrdad_2016_Globecomm_repeatedAuctions}.
In repeated auctions, \acp{UE} can observe patterns in the bidding behavior of others, estimate likely winning bids, and adjust their strategies across rounds to maximize their long-term utility. 
However, relatively little attention has been paid to how users can strategically learn and optimize their bids in these dynamic environments.

\subsection{Related Works}
Another important dimension often overlooked in the existing literature concerns the architectural assumptions underlying the auction process. In many auction-based spectrum allocation studies (e.g.,~\cite{Dong_INFOCOM_2014, Alibraheemi_ACCESS_2024}), a centralized entity is assumed to orchestrate the resource distribution process across all \acp{BS}. Under this assumption, users typically submit a single bid that applies uniformly across BSs, allowing the central controller to select winners and allocate channels optimally. While this simplifies the allocation logic, it abstracts away from the decentralized and heterogeneous nature of practical HetNets. In reality, users often encounter multiple BSs, each with its own traffic load, interference profile, and pricing strategy. In such distributed settings, users must make a two-stage decision: first, select a BS for potential association, and second, determine a bid tailored to that BS's characteristics. This layered decision-making introduces complex strategic considerations -such as load balancing, bid competition, and risk management- that are not captured by conventional single-auction models. These limitations highlight the need to explore distributed auction mechanisms and intelligent user-side decision strategies in more depth.

Recently, \acp{LLM} have demonstrated remarkable abilities in context understanding, probabilistic reasoning, and strategy formulation across diverse domains~\cite{Ismail_2024_Rethinking, sun2025_game_theory_meets_LLMs}. These models can interpret rich interaction histories, infer opponent behavior, and generate adaptive actions aligned with long-term objectives. Such capabilities make them promising candidates for integration into UE decision-making processes, particularly in dynamic wireless environments where budget constraints, user heterogeneity, and competitive bidding interplay. Moreover, the emergence of Agentic-AI frameworks has enabled LLMs to act as autonomous agents capable of negotiation, collaboration, and strategic competition~\cite{Park2023, Ismail_2024_Rethinking}. For instance, in~\cite{mao_2024_Alympics_LLM}, a Microsoft team demonstrated that LLMs could successfully participate in single-shot, first-price auctions, highlighting their potential to internalize game-theoretic reasoning.
Building upon these insights, our work extends the application of LLMs to a more complex and realistic setting -repeated multi-channel spectrum auctions in HetNets- where resource availability, interference, and budget dynamics evolve over time.

\subsection{Contributions}

In this work, we propose a distributed, repeated multi-channel auction framework for HetNets where each BS independently allocates spectrum, and UEs autonomously decide their association and bid values under budget constraints. This decentralized formulation captures realistic congestion scenarios where user demand exceeds available channels, transforming resource allocation into a long-term economic decision rather than a static optimization problem. Within this framework, we investigate LLM-driven UEs that reason over past outcomes and adapt bidding strategies to maximize long-term utility. The heterogeneous HetNet environment -with diverse BS types, interference levels, and cost structures- provides an ideal testbed to evaluate such reasoning capabilities. 
Our contributions are summarized as follows:

\begin{itemize}
    \item We introduce a novel framework for repeated multi-channel spectrum auctions in distributed \acp{HetNet}, where UEs employ LLMs as strategic reasoning agents to guide bidding and association decisions.

    \item Unlike traditional studies that focus solely on auctioneer-side optimization, we analyze how UEs with budget constraints develop bidding strategies in dynamic, multi-BS settings. We explore several strategies, including myopic, greedy and LLM-based reasoning, and examine their impact on long-term utility and channel allocation efficiency.

    \item Through extensive simulations, we analyze the behavior of LLM-based UEs relative to classical bidding strategies. Results show that the LLM-enabled UE discovers adaptive, utility-enhancing behaviors, achieving up to 50\% higher bid precision and 20\% greater channel access frequency compared to both myopic and greedy counterparts~\cite{Mehrdad_2016_Globecomm_repeatedAuctions, athey2006empirical}.
    
\end{itemize}

The rest of the paper is organized as follows. In Section~\ref{sec_system_model} we present our studied system and define the evaluation metrics. In Section~\ref{sec_pb_formula} we derive the BS selection and channel allocation procedures. We present our results in Section~\ref{sec_results} and conclude the paper in Section~\ref{sec_conclusion}.

\section{System Model}\label{sec_system_model}
\subsection{System Description}

\begin{figure}[htbp]
    \centering
    \includegraphics[width=\columnwidth]{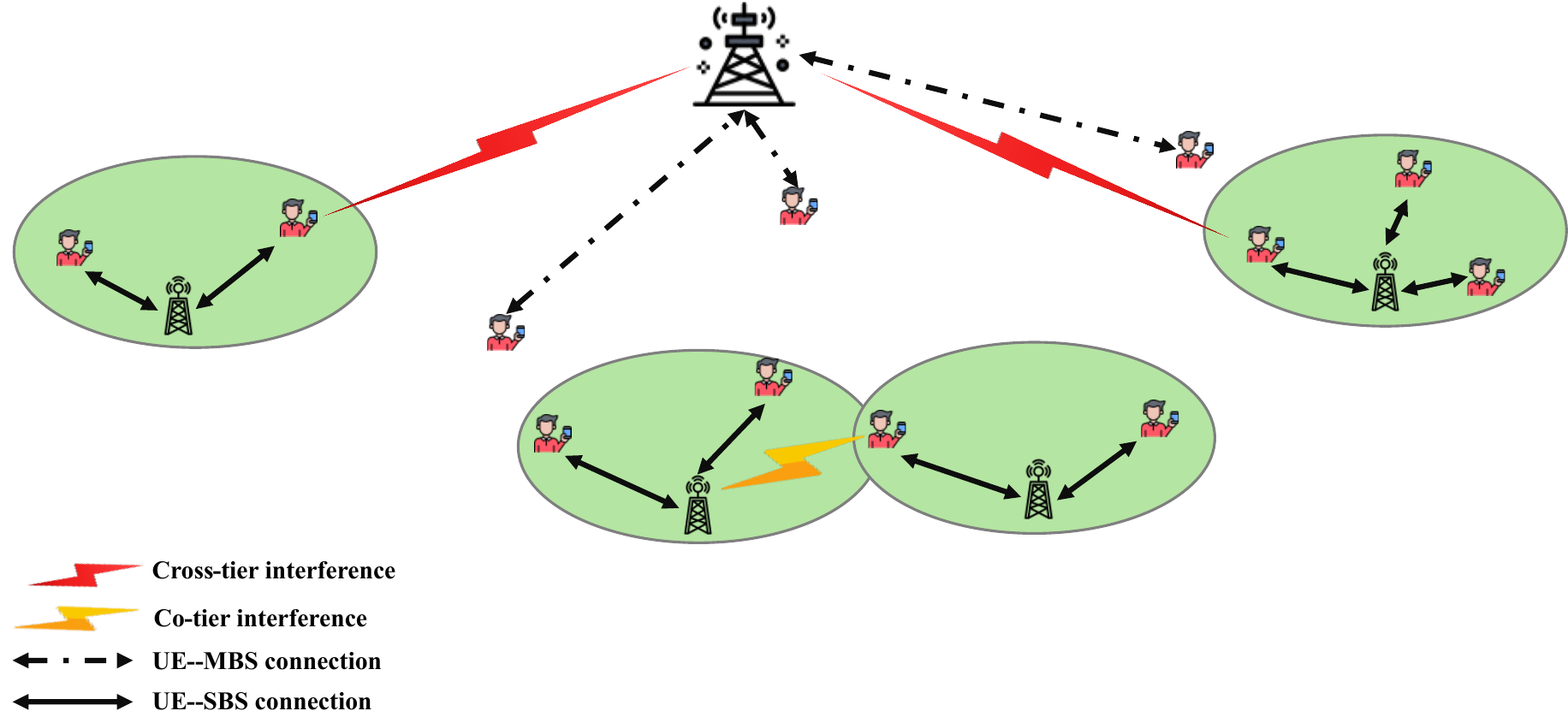}
    \caption{An example of a two-tier HetNet including one macrocell BS and four small BSs.}
    \label{fig:case_1_HetNet}
\end{figure}

Consider a two-tier HetNet comprising a single \ac{MBS} and $M-1$ \acp{SBS} as illustrated in Figure~\ref{fig:case_1_HetNet}. 
$\mathcal{S}$ represents the set of BSs and $\mathcal{C}_s$ denotes the set of sub-channels available to each BS where $s \in \mathcal{S}$. We assume that each BS $s$ transmits over sub-channel $k \in \mathcal{C}_s$ with a fixed power level denoted by $P_{s}$. 
The transmit power of the SBSs is assumed to be equal while that of the MBS is much higher.
Let $\mathcal{U}$ be the set of $N$ UEs located within the service area of the HetNet and $\Psi =\{\psi_1, \psi_2,\dots, \psi_i, \dots, \psi_{N}\} $ denote the set of budgets for each UE. Each UE $i \in \mathcal{U}$ requests a downlink data rate $\overline{R}_i \in \Omega$, where $\Omega$ is a predefined discrete set representing the available service classes (in bits per second)\footnote{Note that, hereinafter, the terms service classes and \ac{QoS} levels are used interchangeably.}.

\subsection{Evaluation Metrics: Data Rate and QoS}

The achievable data rate $R_{s,i}$ for UE $i$ on a sub-channel of BS $s$ is modeled using the Shannon capacity formula as~\cite{ThantZin_2017_TMC}:

\begin{equation}\label{eq_data_rate_k}
R_{s,i} = W \cdot \log_2(1 + \gamma_{s,i}),
\end{equation}

\noindent where $W$ is the bandwidth of a single sub-channel and $\gamma_{s,i}$ is the instantaneous \ac{SINR} at UE $i$, when served by BS $s$.
To model the total data rate achieved by UE $i$, we introduce the following binary decision variables:
\begin{itemize}
    \item $x_{s,i} \in \{0,1\}$: Indicates whether UE $i$ is associated with BS $s$ (i.e., $x_{s,i} = 1$ if UE $i$ is served by BS $s$, otherwise 0).

    \item $y_{s,i}^{(k)} \in \{0,1\}$: Indicates whether a sub-channel $k$ is allocated by BS $s$ to UE $i$.
\end{itemize}

Then, the total data rate $\hat{R}_{s,i}$ achieved by UE $i$ across all its allocated $N_{s,i}$ sub-channels from BS $s$ is given by:

\begin{equation}\label{eq_data_rate_total}
\hat{R}_{s,i} = \sum_{s \in \mathcal{S}} x_{s,i} \sum_{k \in \mathcal{C}_s} y_{s,i}^{(k)} \cdot R_{s,i}.
\end{equation}

Note here that in our system, UE $i$ can be only associated to one BS as in~\cite{ThantZin_2017_TMC}, i.e.,:

\begin{equation}
    \sum_{s \in \mathcal{S}} x_{s,i} \leq 1, \quad \forall i \in \mathcal{U}.
\end{equation}

Additionally, for assignment feasibility, a minimum QoS should be guaranteed by the BS serving UE $i$, which is translated into the following constraint:

\begin{equation}\label{eq_QoS_constraint}
    \hat{R}_{s,i} \geq \overline{R}_i, \quad \forall i \in \mathcal{U}.
\end{equation}

\subsection{Utility Functions}
\subsubsection{BS Utility}

Since electricity consumption constitutes a significant portion of a BS’s operational cost, we model the downlink operational cost primarily in terms of energy expenditure. Specifically, we define the operational cost incurred by BS $s$ when transmitting on a sub-channel as:

\begin{equation}\label{eq_reservation_price}
    r_{s} = \mu_b P_{s},
\end{equation}

\noindent where $\mu_b$ is the unit price of transmission power at BS $s$.
The operational cost defined in~\eqref{eq_reservation_price} is considered as the reservation price (i.e., minimum price) set by the BS for each sub-channel. As the MBS consumes more energy to transmit, its reservation price is considered to be much higher than that of the SBSs.
The utility of BS $s$ from serving UE $i$ at time period $t$ is then defined as the difference between the payment received from UE $i$ at time period $t$ and the operational cost, which we denote formally as:

\begin{equation}\label{eq_utility_BS}
    u_{s\rightarrow i}^{(t)} = N_{s,i} (\pi_{i}^{(t)} - r_{s}),
\end{equation}

\noindent where $\pi_{i}^{(t)}$ is the payment of UE $i$ for each allocated sub-channel at time period $t$.

\subsubsection{UE Utility}
The utility of UE $i$ is defined as the difference between its valuation of all the allocated sub-channels and the payment given to BS $s$. Formally, this is expressed by the following quasilinear preference
function:

{\small \begin{equation}\label{eq_utility_UE}
u_{i\rightarrow s}^{(t)} = \begin{cases} 
N_{s,i} (v_{i\rightarrow s}^{(t)} - \pi_{i}^{(t)}), & \text {if UE $i$ wins},
\\ 0, & \text {otherwise}.
\end{cases}
\end{equation}}

\noindent where $v_{i\rightarrow s}^{(t)}$ is the per-channel valuation of UE $i$ at time period $t$.

A fundamental outcome of any auction is the partition of participants into winners and losers. For wireless users, losing an auction round has a direct behavioral consequence: 
since connectivity is required continuously to sustain the target QoE, each missed opportunity 
increases the urgency of securing resources in subsequent rounds. This urgency translates into a dynamic adjustment of the user’s valuation of spectrum access. We capture this effect by modeling the valuation of UE~$i$ for association with BS~$s$ at round~$t$ through a time-dependent quasi-linear utility function: 

\begin{equation}
    v_{i \rightarrow s}^{(t)} = \alpha_i^{(t)} \, R_{s,i},
\end{equation}

\noindent where $R_{s,i}$ is the achievable data rate defined in~\eqref{eq_data_rate_k}, and 
$\alpha_i^{(t)}$ is a user-specific scaling parameter that evolves with time. 
The parameter $\alpha_i^{(t)}$ reflects how strongly throughput translates into perceived utility, capturing both the intrinsic service requirements of the user (e.g., real-time vs. elastic traffic) and the urgency induced by repeated failures to secure a channel.
We model the time evolution of the per-user scaling parameter \(\alpha_i^{(t)}\) as a function of the user's history of missed allocations. Let $h_i(t)$ denote the number of consecutive rounds up to $t$ in which UE $i$ failed to obtain a channel.
Let \(\alpha_i^{(0)}\) denote the baseline scaling and \(\bar\alpha_i\) a user-specific upper bound (urgency ceiling). The update rule is then defined as:
\begin{equation}
    \alpha_i^{(t)} = \min\Big\{\bar\alpha_i,\ \alpha_i^{(0)} + \varepsilon_i \, h_i(t)\Big\},
\end{equation}

\noindent where $\varepsilon_i = (\bar\alpha_i-\alpha_i^{(0)})/H_{\text{sat}}$ and $H_{\text{sat}}$ is the number of failures to reach saturation (e.g., $H_{\text{sat}}=5)$.
This formulation indicates that each consecutive failure increases urgency by a fixed increment $\varepsilon_i$, until capped at $\bar\alpha_i$\footnote{Note that other quasi-linear variants (e.g., logarithmic or sigmoidal) are straightforward to incorporate within the same framework and yield qualitatively similar dynamics.}.
By adapting this formulation, user valuations become endogenously time-varying rather than fixed constants, linking auction outcomes to future bidding behavior.

\section{Problem Formulation: Auction-based Distributed Channel Allocation}\label{sec_pb_formula}

In this section, we formalize the problem setting and the decision-making processes of both UEs and BSs. Existing works typically assume that BS selection is centrally optimized by the MBS under strict rate and power constraints, often resulting in NP-hard formulations that are relaxed and solved using centralized heuristics~\cite{Yang_2013_COMST_AuctionTheory}. While effective in static settings, such approaches limit scalability and responsiveness in dynamic HetNets. In contrast, we focus on the more realistic scenario of congestion, where user requests exceed the number of available channels, necessitating competitive allocation. To address this, we adopt a UE-driven association model in which each UE initially requests service from the SBS with the highest SINR and only resorts to the MBS if unsuccessful.

\subsection{BS Selection Procedure (UE-Side Decision)}

The UE-based association changes the decision structure compared to existing works: instead of being forced by feasibility (e.g., power constraint), the UE must anticipate \emph{which association request is most likely to succeed and which option maximizes its expected utility subject to winning.} 
A naive BS selection rule would be to always associate with the SBS providing the highest SINR. While this maximizes signal quality, it also concentrates competition, driving up clearing prices and reducing the likelihood of winning. On the other hand, consistently choosing the MBS increases winning probability but incurs higher payments, leading to rapid budget depletion and inefficient connectivity. These trade-offs highlight the need for more strategic and context-aware decision-making, where UEs weigh both expected utility and winning probability. To this end, we investigate three approaches for UE-side decision-making in BS and bid selection: myopic~\cite{Mehrdad_2016_Globecomm_repeatedAuctions}, greedy~\cite{athey2006empirical, baltaoglu2017online}, and LLM-based strategies.

\subsubsection{\textbf{Myopic Truthful Strategy}}
The UE ignores cross-BS tradeoffs and simply bids its true valuation at the BS with the highest achievable rate as defined in~\eqref{eq_data_rate_total}. 



\subsubsection{\textbf{Greedy Bayesian Strategy}}


The UE evaluates both SBS and MBS in the current round, comparing expected utilities (defined below), and chooses the pair of selected BS and bid value $(s^\star,b^\star)$ that maximizes the instantaneous expected utility.  
Formally, the expected utility for UE $i$ from association with BS $s$ at time period $t$ is defined as:
\begin{equation}\label{eq:expected_util}
    \hat{U}_i(b,s)\;=\;\Pr(\text{win}\mid b)\cdot\Big(v_{i,s}^{(t)} - \mathbb{E}[p_s]^{(t)}\Big),
\end{equation}

\noindent where $\Pr(\text{win}\mid b)$ is the winning probability at BS $s$ with bidding $b$ and  $\mathbb{E}[p_s]^{(t)}$ denotes the expected payment at BS $s$.

\paragraph{Winning probability and expected payment}
UE \(i\) wins a channel at BS \(s\) with bid \(b\) if fewer than \(\mathcal{C}_s\) competitors submit bids strictly greater than \(b\). Let $N_s$ be the number of competing UEs at BS $s$ and \(B_{-i} = \{B_j\}_{j\neq i}\) be the random bids of the other \(m=N_s-1\) competitors. The number of competitors whose bids exceed \(b\) follows a binomial distribution with parameters \(m\) and \(p_{\text{lose}}(b)=1-F_s(b)\) where $F_s(b)\;=\;\Pr\big(B_j \le b\big)$ is the CDF of the competitors bids at BS $s$. Hence, the win probability is\footnote{Here we assumes that all competing bids are independent and identically distributed (i.i.d.) according to the empirical CDF $F_s(\cdot)$ of BS~$s$, which is common in auction theory practices~\cite{krishna2009auction}.
}:

\begin{equation}\label{eq:win_prob}
\Pr(\text{win}\mid b)
=\sum_{j=0}^{\mathcal{C}_s-1} \binom{m}{j} \big(1-F_s(b)\big)^j \, \big(F_s(b)\big)^{\,m-j}.
\end{equation}


This expression integrates over all events where exactly \(j\) competitors outbid user \(i\), and user \(i\) still ranks among the top \(\mathcal{C}_s\) bidders.
Here we sum only up to $\mathcal{C}_s-1$ because the UE can tolerate at most $\mathcal{C}_s-1$ competitors outbidding it and still winning one of the $\mathcal{C}_s$ available sub-channels.
However, unlike theoretical models where \(F_s\) is known, a UE in practice must infer it from broadcast information. 
In practice, the BS may either (i) broadcast a full vector of winning payments, or (ii) broadcast only the marginal clearing price (i.e., the highest losing bid). 
The former provides complete information but incurs significantly higher signaling overhead, as the BS must communicate one payment per allocated channel in every round. By contrast, the latter requires broadcasting only one value -the marginal clearing price (the 
$C_s$-th order statistic)- which greatly reduces control-channel load and still provides sufficient statistics for UEs to estimate the competitiveness of the market. In this work, we adopt the single-clearing-price model to keep the feedback lightweight, though extending to per-winner payment broadcasts is conceptually straightforward.
From the sequence of clearing prices up-to time period $t$: \(\{p_s^{(1)},p_s^{(2)},\dots,p_s^{(t)}\}\), UE \(i\) builds an empirical distribution~\cite{athey2006empirical}:
\begin{equation}
    \hat{F}_s^{(t)}(x) \;=\; \frac{1}{t}\sum_{\tau=1}^t \mathbf{1}\{p_s^{(\tau)} \le x\}.
\end{equation}

\noindent where $\mathbf{1}\{\cdot\}$ is the indicator function. 
This estimated CDF $\hat{F}_s$ feeds directly into Eq.~\eqref{eq:win_prob}, allowing each UE to compute its personalized probability of winning at bid \(b\).
From this empirical distribution, the expected marginal payment is directly estimated as:
\begin{equation}
    \mathbb{E}[p_s]^{(t)} \;\approx\; \frac{1}{t}\sum_{\tau=1}^{t} p_s^{(\tau)}.
\end{equation}

\paragraph{Association decision}
UE \(i\) constructs a candidate bid set \(\mathcal{B}_i\) as a discrete grid around two anchors: its own valuation \(v_{i,s}^{(t)}\) and the last observed minimum clearing price \(\hat{p}_s^{(t)}\). This ensures exploration of both aggressive (above \(\hat{p}_s^{(t)}\)) and conservative (near \(v_{i,s}^{(t)}\)) bids, while maintaining budget feasibility.
Finally, the UE solves:
\begin{equation}
    (s^\star,b^\star)\;=\;\arg\max_{s\in\mathcal S,\ b\in\mathcal B_i} \hat{U}_i(b,s).
\end{equation}

\subsubsection{\textbf{LLM-based Strategy}}

To support advanced decision-making, recent studies have highlighted the potential of \acp{LLM} as lightweight yet powerful solvers for complex optimization tasks when guided by carefully designed prompts~\cite{yang2023large}. Motivated by this, we explore an \emph{LLM-driven bidding policy}, where each UE relies on an embedded LLM agent to determine both the preferred BS and the corresponding bid to submit. Specifically, for each candidate BS $s \in \mathcal{S}'$, the UE assembles a structured prompt that summarizes its current requirements. This prompt is processed by the internal LLM module, which outputs a decision tuple $(s^*, b^*, \mathcal{E})$. Here, $s^*$ denotes the chosen BS, $b^*$ is the recommended bid value, and $\mathcal{E}$ is a natural language explanation detailing the reasoning behind the suggestion.  
An illustrative example of such an LLM-based prompt, along with a representative response format, is depicted in Fig.~\ref{fig:llm_prompt}.

\begin{figure}[t]
\begin{tcolorbox}[colback=brown!10, colframe=brown!50, width=1\linewidth, boxrule=0.5pt]
\centering
\fbox{
\begin{minipage}{0.95\linewidth}
\tiny
\textbf{LLM Prompt Template:}

\texttt{
Given the following network and economic context: \\
- Your valuations for BSs in $\mathcal{S}'$ \\
- Your budget: $\beta_i$ \\
- Number of sub-channels required for each BS: $N_{s,i}$ \\
Please analyze and provide: \\
1. The optimal BS $s^*$ to associate with, provided it has the highest expected utility. \\
2. Recommended bid value for BS $b^*$. \\
3. A brief explanation of your reasoning.
}

\textbf{Expected Response Format:}

\texttt{Selected BS and bid value: [value] \\
Explanation: "[Short textual reasoning]"
}
\end{minipage}
}
\end{tcolorbox}
\caption{Template of the LLM prompt and expected response structure used for strategic BS selection and bidding.}
\label{fig:llm_prompt}
\end{figure}

\subsection{Channel Allocation Procedure (BS-Side decision)}

After selecting a serving BS, each UE $i$ submits a single association request within a given time slot. This request specifies both the desired number of sub-channels and the price it is willing to pay. 
Furthermore, the resource allocation at each BS is carried out through an auction mechanism that operates periodically over an allocation window of length $T_c$. We assume $T_c$ coincides with the coherence time of the channel, i.e., the duration for which the channel state remains approximately constant. This design allows the system to re-optimize allocation at a timescale consistent with physical-layer dynamics, thereby accommodating traffic fluctuations and varying channel conditions while containing signaling overhead\footnote{Note that while current LLMs exhibit higher inference latency than heuristic methods, this limitation is expected to diminish with the advent of lightweight and optimized future LLM models.}. 
An entrance fee is considered for each UE that chooses to participate in the auction regardless of the auction outcome.
Here, the BSs adopt the widely studied multi-unit \ac{VCG} auction~\cite{krishna2009auction}. In VCG, each winning bidder pays the marginal externality it imposes on others. This mechanism is incentive-compatible in a single-shot setting without budget limits, where truthful bidding forms a Nash equilibrium. 

\section{Experimental Analysis}\label{sec_results}

To evaluate the efficacy of the proposed LLM-based bidding strategy, we conduct a series of simulations where multiple UEs compete for sub-channel allocations through repeated VCG auctions.
Performance is assessed in terms of average utility, bid-winning frequency, and channel access rates across varying auction horizons and system parameters.
We consider a two-tier HetNet composed of one MBS and two SBSs. Unless otherwise specified, the number of available sub-channels per BS is set to $C_s=4$, and the number of UEs is set to $40$. Each UE is equipped with a finite budget $\psi_i=15~\textit{units}$ that is not refilled over the simulation horizon. 
For the LLM-based bidder, we employ the \texttt{gpt-5-mini} model, chosen for its favorable trade-off between reasoning capability and low inference latency.

\subsection{LLM and Greedy UEs Competing in a Myopic Population}

\begin{figure}[htbp]
    \centering
    \includegraphics[width=0.8\columnwidth,height=3.1cm]{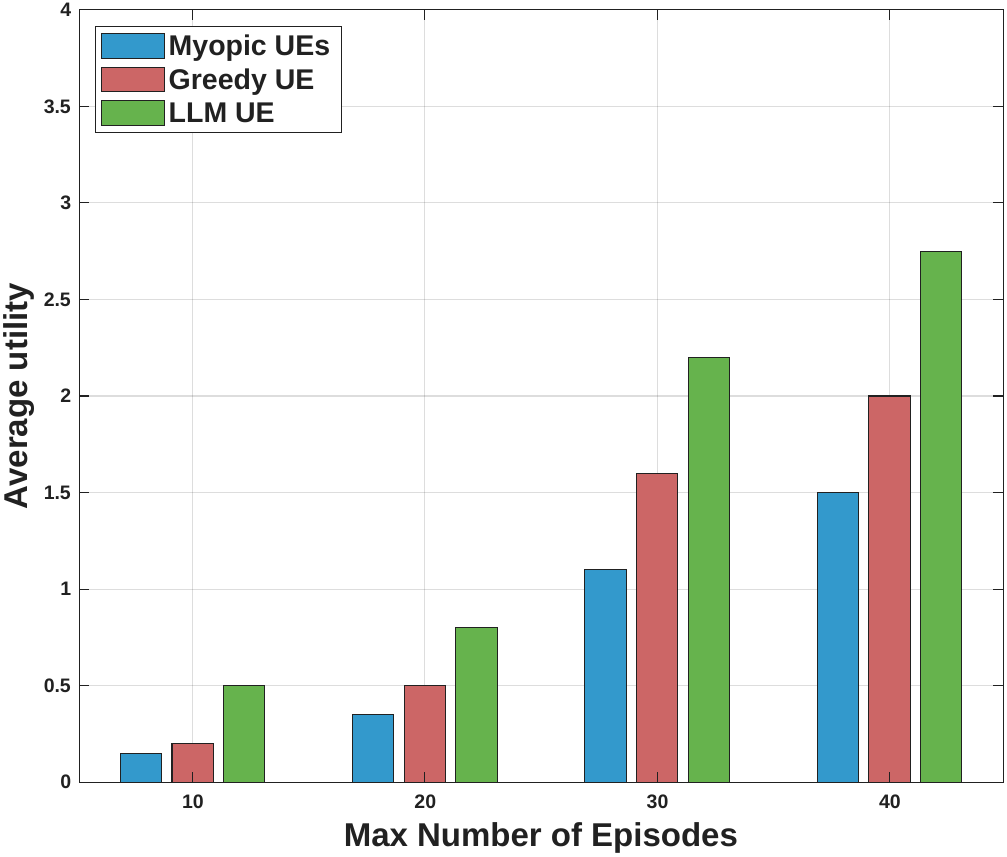}
    \caption{Average utility for different bidding strategies.}
    \label{fig:vcg_avg_utility}
\end{figure}

We first consider a scenario with myopic majority where only one UE uses LLM-based reasoning and one UE uses the greedy method while the remaining UEs use the myopic strategy.
The results in Figure~\ref{fig:vcg_avg_utility} demonstrate a clear increase in average utility (i.e., the mean utility over multiple simulation runs -computed per user for the LLM-based and the greedy UEs, and averaged across all users for the myopic UEs) for all bidding strategies as the maximum number of episodes grows. This trend arises because, in early-ending auctions, many UEs still retain unused budget, preventing them from fully realizing their potential gains. 
When comparing strategies, the LLM-based UE consistently outperforms both myopic and greedy UEs, and the performance gap widens with longer horizons. The superiority of the LLM stems from its ability to reason strategically about budget allocation across future rounds. Unlike myopic and greedy strategies, which bid aggressively without foresight, the LLM anticipates lower competition in later episodes. By conserving budget early on, it is able to exploit reduced bidding intensity in subsequent rounds, where VCG payments are typically lower due to weaker competition~\cite{krishna2009auction}. This budget-aware reasoning enables the LLM to achieve substantially higher utilities, particularly in longer auction horizons, highlighting its effectiveness in repeated spectrum allocation settings.

\begin{figure}[ht] 
     \centering
     \begin{subfigure}[b]{0.24\textwidth}
         \centering         \includegraphics[width=\textwidth,height=3.1cm]{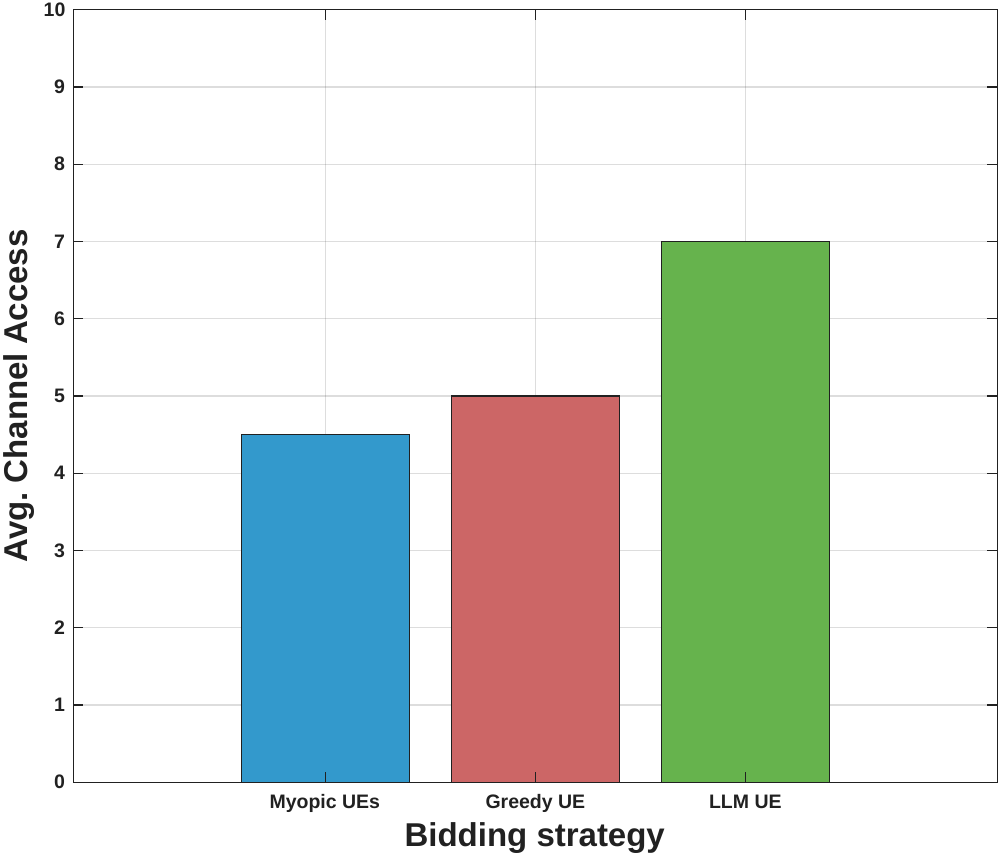}
         \caption{}
         \label{fig:vcg_avg_channel_acc}
     \end{subfigure}
     \begin{subfigure}[b]{0.24\textwidth}
         \centering         \includegraphics[width=\textwidth,height=3.1cm]{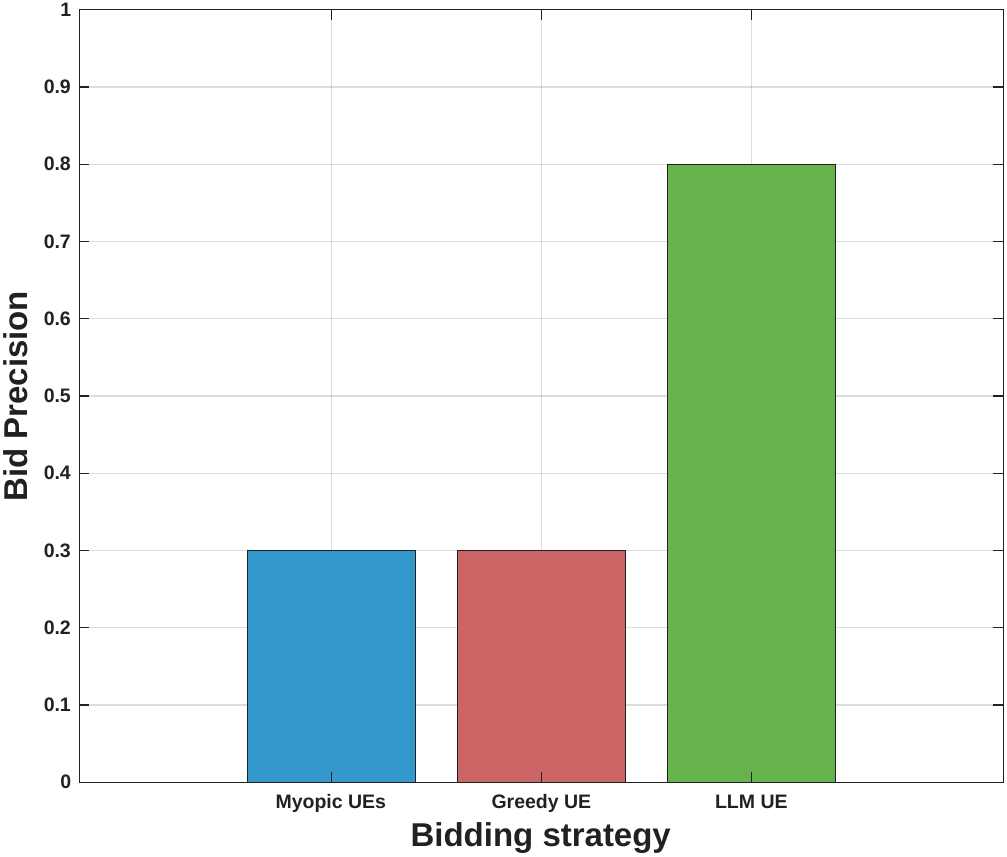}
         \caption{}
         \label{fig:vcg_bid_precision}
     \end{subfigure}
     \caption{(a) average number of channel access and (b) bid precision for different bidding strategies.
        }        \label{fig:vcg_avg_channel_acc_bid_precision}
\end{figure}

To further understand why the LLM is able to outperform other methods, we plot in Figure~\ref{fig:vcg_avg_channel_acc_bid_precision} the average number of channel access and bid precision for the different bidding strategies. The bid precision is defined as the proportion of bids that resulted in a win ((analogous to precision in classification tasks in machine learning).
The LLM bidder consistently secures a greater number of sub-channels and bid precision compared to both myopic and greedy UEs. 
This high bid precision of the LLM UE indicates that the UE bids with greater confidence, successfully acquiring the resources it requests, whereas myopic and greedy UEs must bid in every round regardless of their chances of winning. By leveraging this advantage, the LLM can bid selectively—focusing only on situations where allocation is most beneficial, such as urgent demand or critical resource requirements—thereby improving overall efficiency and utility.

This performance gain arises from the LLM’s reasoning capabilities, which enables it to strategically allocate its budget across auction rounds. By conserving resources in early rounds and targeting high-probability winning opportunities, the LLM effectively leverages accumulated savings -reflected in the increased utility shown in Figure~\ref{fig:vcg_avg_channel_acc}- to acquire approximately $15\%$ more sub-channels than the other users. 
Furthermore, the high bid precision of the LLM UE help reduce unnecessary participation fees, giving access to more resources compared to the other strategies.
This illustrates how budget-aware, LLM-based foresighted bidding can translate into tangible resource advantages in repeated auction settings.

\subsection{Single LLM Against a Greedy Population}

\begin{figure}[ht] 
     \centering
     \begin{subfigure}[b]{0.24\textwidth}
         \centering         \includegraphics[width=\textwidth,height=3.2cm]{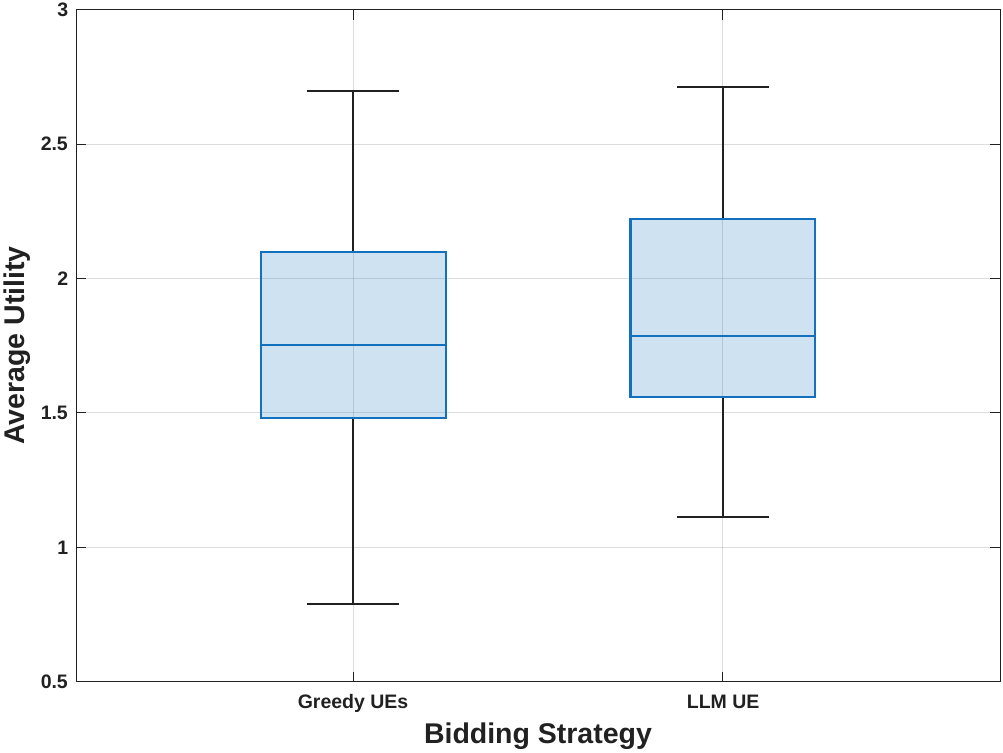}
         \caption{}
         \label{fig:vcg_dynamics_greedy_population_utility}
     \end{subfigure}
     \begin{subfigure}[b]{0.24\textwidth}
         \centering         \includegraphics[width=\textwidth,height=3.2cm]{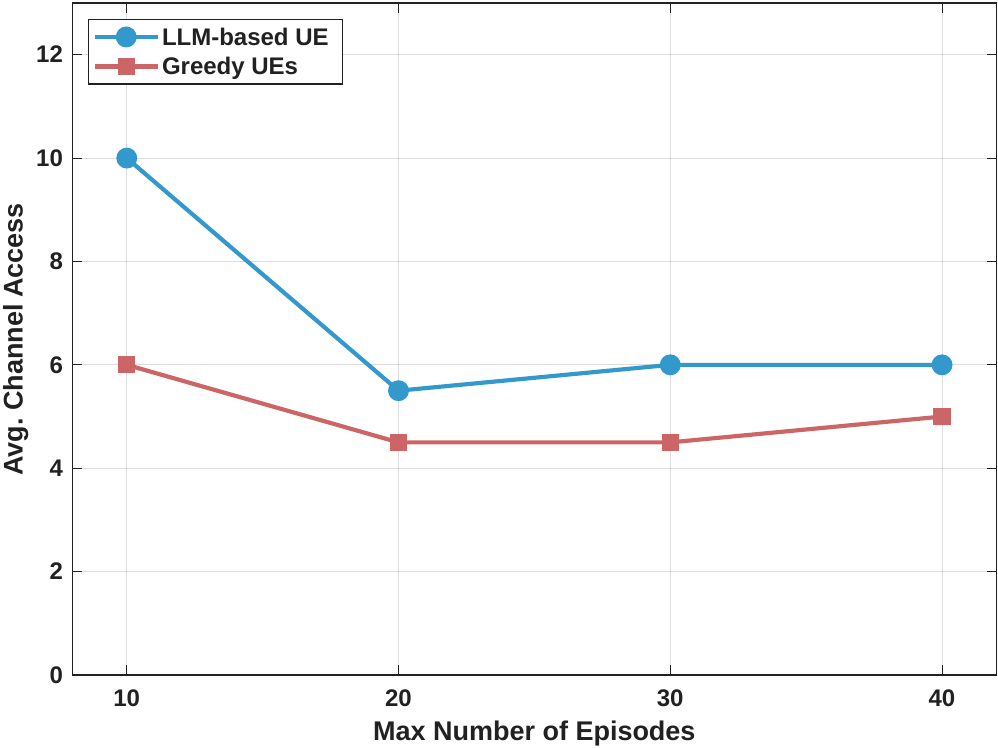}
         \caption{}
         \label{fig:vcg_dynamics_greedy_population_access}
     \end{subfigure}
     \caption{(a) utility variability and (b) average number of channel access for the LLM UE and the other greedy UEs in greedy majority population.
        }        \label{fig:vcg_dynamics_greedy_population}
\end{figure}

Building on the earlier results where the LLM consistently outperformed both myopic and greedy strategies, we now examine whether this advantage persists in a more competitive setting where the majority of UEs adopt a greedy bidding strategy.
As observed from Figure~\ref{fig:vcg_dynamics_greedy_population_utility}, the relative performance shifts: the LLM no longer always outperforms its greedy competitors. Instead, its performance converges toward the top-performing greedy UEs. In some instances, greedy users surpass the LLM in terms of average utility. This phenomenon arises because a predominantly greedy population generates more volatile and competitive clearing prices, reducing the stability and predictability that the LLM exploited in the first scenario.

However, despite the reduced utility margin, the LLM-based UE continues to dominate in terms of channel access frequency, as shown in Figure~\ref{fig:vcg_dynamics_greedy_population_access}, particularly in short-horizon simulations. For a small maximum number of episodes, the LLM secures significantly more sub-channel allocations than its greedy counterparts. This advantage arises from its bid precision and reasoning capability, which allow it to selectively participate only in rounds with a high probability of success, thereby conserving budget and avoiding unnecessary auction entrance costs. Such behavior is particularly advantageous in HetNets, where fluctuating interference and user density demand adaptive decision-making. This suggests that when a UE urgently needs to secure sub-channels within a limited number of rounds (e.g., under high-traffic or delay-sensitive conditions), LLM-guided bidding provides a clear advantage. Although the gap narrows over longer horizons -as all UEs gradually deplete their budgets- the LLM-based UE still maintains roughly a 10\% improvement in spectrum access efficiency.

\section{Conclusion}\label{sec_conclusion}
This paper presented a novel exploration of LLM-augmented bidding strategies in repeated multi-channel spectrum auctions for \acp{HetNet}. We proposed a distributed architecture where each BS independently runs its own auction, and UE agents strategically select both the BS to associate with and the corresponding bid value. By integrating reasoning capabilities into UEs' decision-making, our framework bridges auction theory with emerging AI-driven autonomy in next-generation wireless systems. Simulation results revealed that the LLM-based UE achieves higher bid precision and more frequent channel access compared to classical myopic and greedy bidders, particularly in short-horizon and congested auction settings.
Even when the competitive dynamics reduce the utility margin, the LLM agent maintains a consistent advantage in access efficiency, highlighting its ability to adaptively manage budget and participation. 
These findings demonstrate the potential of reasoning-driven bidding as a foundation for future adaptive spectrum management mechanisms. While current LLMs remain computationally demanding for deployment on mobile devices, ongoing advancements in lightweight models and edge inference suggest that such intelligent bidding systems could become practical components of future mobile networks.

\bibliographystyle{IEEEtran}
\bibliography{ref}

\end{document}